# Electron-Hole Separation in Ferroelectric Oxides for Efficient Photovoltaic Responses

Donghoon Kim,[1] Hyeon Han,[1] June Ho Lee,[1] Jeffrey C. Grossman,[2] Donghun Kim,[3*] Hyun Myung Jang[1*]

[1] Department of Materials Science and Engineering, and Division of Advanced Materials Science, Pohang University of Science and Technology (POSTECH), Pohang 37673, Republic of Korea

[2] Department of Materials Science and Engineering, Massachusetts Institute of Technology (MIT), Cambridge, MA 02139, United States

[3] Computational Science Research Center, Korea Institute of Science and Technology (KIST), Seoul 02792, Republic of Korea

[*]Correspondence to: donghun@kist.re.kr (D.K.); hmjang@postech.ac.kr (H.M.J.)

**Abstract**

Despite their potential to exceed the theoretical Shockley-Queisser limit, ferroelectric photovoltaics (FPVs) have performed inefficiently due to their extremely low photocurrents. Incorporating $Bi_2FeCrO_6$ (BFCO) as the light absorber in FPVs has recently led to impressively high and record photocurrents [Nechache *et al. Nature Photon.* **2015**, *9*, 61], reviving the FPV field. However, our understanding of this remarkable phenomenon is far from satisfactory. Here, we use first-principles calculations to determine that such excellent performance mainly lies in the efficient separation of electron-hole (e-h) pairs. We show that photoexcited electrons and holes in BFCO are spatially separated on the Fe and Cr sites, respectively. This separation is much more pronounced in disordered BFCO phases, which show exceptional PV responses. We further set out to design a strategy for next-generation FPVs, not limited to BFCO, by exploring 44 additional Bi-based double-perovskite oxides. We suggest 9 novel active-layer materials that can offer strong e-h separations and a desired band gap energy for application in FPVs. Our work indicates that charge separation is the most important issue to be addressed for FPVs to compete with conventional devices.

Ferroelectrics have long garnered attention in light-to-electricity conversion devices owing to their anomalously high photovoltages[1-7], coupled with their reversibly switchable photocurrents[8-9]. However, ferroelectric photovoltaics (FPVs) suffer from extremely low photocurrents (on the order of $\mu A/cm^2$ under 1 sun illumination), which has mainly been attributed to the wide band gap energy (>2.5eV) of the active-layer materials (e.g., $BiFeO_3$ (BFO) and $Pb(Zr_xTi_{1-x})O_3$ (PZT))[10-11]. Tremendous effort has been spent to overcome this challenge, utilizing techniques such as band gap tuning[5,12], domain-structure manipulation[2,13], and multi-junction stacking[14,15]. Unfortunately, the power conversion efficiencies (PCEs) of FPVs have until recently remained too low ($\ll$1%) to be utilized in practical PV applications[6].

In this regard, the recent work of Nechache *et al.*, in which a new light-absorbing material, namely, $Bi_2FeCrO_6$ (BFCO) was employed in a double-perovskite structure, is highly noteworthy[16]. In 2005, P. Baettig *et al.* predicted from first-principles calculations that multiferroic BFCO simultaneously possesses large spontaneous polarization and magnetization above room temperature[17,18]. Motivated by this *ab initio* prediction, Nechache *et al.* synthesized BFCO-based multi-layers using the pulsed laser deposition (PLD) technique[19-21]. Incorporating BFCO as the light absorber remarkably improved the FPV device performance: photocurrents on the order of tens of $mA/cm^2$ and a record PCE of 8.1% were achieved[16]. This work is indeed a breakthrough, as it solved the biggest problem (i.e., low photocurrent) that FPVs have faced since their inception.

Herein, we reveal the main origin of the phenomenal performance of BFCO-based FPVs to be efficient electron-hole (e-h) separation. Utilizing *ab initio* density functional theory (DFT) calculations, we show that the photoexcited e-h pairs in BFCO are spatially separated on Fe and Cr sites, respectively, irrespective of the configuration (spatial arrangement) of the B-site cations, leading to low charge recombination rates. This is in sharp contrast to the tetragonal BFO system, in which both electrons and holes occupy the same sites (i.e., Bi orbitals), which increases the e-h recombination rate. We further find that the positional disordering of B-site cations in BFCO substantially enhances the degree of e-h separation, which is critical to understanding the observed exceptional PV responses. With this knowledge, we then establish a design strategy for next-generation FPV materials that can outperform BFCO. We explore 44 additional Bi-based ferroelectric oxides in a double-perovskite structure and suggest 9 novel materials that can potentially offer an efficient e-h separation route and a desired band gap for application in FPVs.

The observed remarkable performance of BFCO-based FPVs, compared with that of prototypical BFO-based FPVs, was previously thought to benefit mainly from band gap reduction and the resultant increase in solar absorption by BFCO[16]. Though it is true that the enhanced solar absorption is an important factor, the magnitude of its contribution has not been quantified. Here, we clarify this point by experimentally measuring the PV parameters of BFO-based cells and comparing these values with those of BFCO-based cells under the same conditions (i.e., active-layer thickness (100 nm), substrate ($SrTiO_3$), electrodes ($InSnO_3$, $SrRuO_3$)). Under AM1.5G illumination, the short-circuit current ($J_{sc}$) of the reference thin-film BFO cell is ~11.6 μA·cm$^{-2}$, while the $J_{sc}$ of the BFCO-based cell is 11.7 mA·cm$^{-2}$. Note the *3-orders-of-magnitude* difference in the $J_{sc}$ values (see Table S1 and Fig. S1 in the Supporting Information for detailed comparisons). Next, we compare the solar absorption (**A**) of the BFO and BFCO films using Eq. (1)

$$\mathbf{A}\ (\%) = \left(1 - \frac{\int_0^{E_g} S_E dE}{\int_0^{\infty} S_E dE} - \frac{\int_{E_g}^{\infty} \exp(-\alpha d) S_E dE}{\int_0^{\infty} S_E dE}\right) \times 100 \qquad (1),$$

where $S_E$ is the solar irradiance as a function of photon energy *E*, α is the *E*-dependent absorption coefficient, and d denotes the film thickness. As shown in Fig. 1a, the BFCO film exhibits a relatively lower absorption onset energy, enabling the absorption of sunlight in the visible range. As a result, the BFCO layer can absorb approximately 4.5 times more sunlight than the BFO layer at the same film thickness of 100 nm (Fig. 1b). These comparisons indicate that increased solar absorption in the BFCO film (only a 4.5-fold enhancement) cannot be the primary factor responsible for the *3-orders-of-magnitude* change in the $J_{sc}$. Thus, the main cause of the remarkably enhanced $J_{sc}$ in BFCO is an open question.

To reveal the origin of this exceptional PV response, we investigated the electronic structure of BFCO. Modeling BFCO requires a large set of computations due to the complexity of the atomic and magnetic orderings. Fig. 1c shows the BFCO model structures used in our simulations. Because the three-dimensional configuration of the B-site cations (i.e., Fe and Cr) in BFCO varies significantly with deposition conditions[16], we considered 10 possible B-site cation configurations in 40-atom unit cells (8 formula units) while maintaining the Fe:Cr ratio at 1:1. These include 9 disordered structures (labeled $d_1$-$d_9$) and the remaining ordered structure in which the Fe and Cr planes alternate along the [111] direction. In addition to the atomic configurations, magnetic orderings were also considered in the present work (inset of Fig. 1d), as these orderings can be varied through thermal excitation[22] or pressure-induced transitions[23]. As BFCO is known to have antiferromagnetic (AFM) ordering, we considered 3 different types of AFM orderings, namely, A, C and G type, for each atomic configuration.

In Fig. 1d, the computed band gap ($E_g$) is plotted as a function of the corresponding Kohn-Sham (K-S) energy for all possible B-site cation configurations. For ordered BFCO, the K-S energy is nearly independent of the spin state: the A-, C-, and G-AFM orderings are nearly degenerate in energy. On the other hand, for the disordered BFCO structures, the C-AFM and G-AFM orderings generally result in lower K-S energies than the A-AFM case[23]. Due to the relative instability, the A-AFM ordering is unlikely to appear in experimentally prepared disordered BFCO samples and thus is not of interest in the present work. The band gap of ordered BFCO has been experimentally measured to be ~1.6 eV[16], and our calculations reproduce this value well (open symbols in Fig. 1d): 1.65 eV (A-AFM), 1.64 eV (C-AFM), and 1.43 eV (G-AFM). We discovered a very strong correlation between $E_g$ and the K-S energy (with a Pearson correlation coefficient $\rho$ of −0.89), which indicates that stable BFCO structures (mostly in C- and G-AFM orderings) exhibit larger $E_g$ values, up to 1.92 eV. This agrees well with the experimental trend in which disordered BFCO possesses a relatively wider $E_g$ (up to 2.4 eV) than ordered BFCO[16], although our calculations underestimate the magnitude. The analysis of the $E_g$ and K-S energies of the BFCO models confirms that C- and G-AFM orderings well represent the spin states in the disordered BFCO phases observed experimentally[16].

As our model captured the key characteristics of BFCO, we went on to calculate the electronic structures of BFCO and compared them with the tetragonal BFO system. For detailed analysis, we focus only on three selected cases: (a) the BFO structure with C-AFM ordering, (b) ordered BFCO with C-AFM ordering, and (c) the "$d_1$" structure with C-AFM ordering (as a representative of disordered BFCO). Fig. 2a-c illustrates the orbital-resolved band structures and the corresponding density of states (DOS) of these three selected cases. For the tetragonal BFO (Fig. 2a), the edge of the highest valence band (VB) at the $\Gamma$-point is predominantly contributed by Bi 6$s$–O 2$p$ hybrid states, and the lowest conduction band (CB) at the same $\Gamma$-point mainly consists of Bi 6$p$ states. Note that, unlike rhombohedral BFO[24] in which the CB is predominantly composed of Fe 3d states, Bi orbitals mainly contribute to the edges of both the VB and CB states[25] in tetragonal BFO. In contrast, the orbital constitutions near band edges in the BFCO cases (Fig. 2b, 2c) completely differ from those in BFO. For both the ordered and disordered BFCO case, broad Cr 3$d$ ($t_{2g}$)–O 2$p$ hybrid states dominate the top of the VB (at the **A**-point in Fig. 2b and at the **M**-point in Fig. 2c), and relatively narrower Fe 3$d$ ($t_{2g}$) states form the bottom of the CB (at the **Z**-point in Fig. 2b and at the **M**-point in Fig. 2c). For visual confirmation, the computed charge densities of the conduction band minimum (CBM) and valence band maximum (VBM) states of the three different cases are shown in Fig. 2d-f. For BFO (Fig. 2d), both the electron and hole density are located at the Bi sites. In contrast, both the ordered and disordered BFCO (Fig. 2e, 2f) show a distinct spatial separation of the electron and hole densities: the CBM and VBM states are mainly localized at the Fe and Cr sites, respectively. Such spatial separation likely leads to easier separation of the photoexcited e-h pairs in the double-perovskite BFCO. Although only selected cases are shown in the main text, the band structures and DOS of all other atomic and magnetic configurations were found to be very similar (available in Fig. S2 in the Supporting Information), which strengthens our main claims.

We then examined the enhanced e-h recombination lifetime ($\tau$) in BFCO over that of the reference BFO system. Based on Fermi's golden rule, the e-h recombination rate ($1/\tau$) can be written as[26]

$$\frac{1}{\tau} = \frac{4}{3}\frac{\alpha\omega n}{m^2 c^2}|\langle e|p|h\rangle|^2 \tag{2}$$

where $\omega$ is the photon frequency, $n$ is the refractive index, $\alpha$ is the fine-structure constant, and $\langle e|p|h\rangle$ is the momentum matrix element between the electron and hole states. Under the Schrödinger picture based on the electric-dipole approximation[27], the momentum matrix element becomes $+im\omega\langle e|r|h\rangle$. In tetragonal BFO, the VBM mostly consists of the Bi 6$s$ orbital (even parity) at the Bi-ion site, while the CBM mostly consists of the 6$p$ orbital (odd parity) at the same Bi-ion site. Thus, the effective matrix element of the VBM-CBM interband transition, $\langle e|r|h\rangle$, becomes an even-parity function, the integration of which will survive as a non-zero value. In contrast, in BFCO, the matrix element vanishes to zero since the Fe (CBM) and Cr (VBM) ion sites are spatially separated by an oxygen ion, i.e., Fe-O-Cr. Even if the two ions were intimately (hypothetically) close to one another, a symmetry argument would further prohibit a non-zero matrix element since $\langle e|r|h\rangle = \langle\psi_{CBM}|r|\psi_{VBM}\rangle = \langle\psi_{t_{2g}(Fe)}|r|\psi_{t_{2g}(Cr)}\rangle = \int even \otimes odd \otimes even\, dv = 0$. Then, Eq. (2) predicts that the lifetime ratio asymptotically approaches infinity, namely, $\frac{\tau_{BFCO}}{\tau_{BFO}} \to \infty$. Thus, this comparison clearly shows that double-perovskite BFCO would have a much longer e-h recombination lifetime than single-perovskite BFO.

To describe more realistic and disordered BFCOs, it is necessary to investigate the charge-separation phenomenon *in much larger scales*. In fact, the simulation cells (local $d_1$-$d_9$) in Fig. 1c are not fully disordered due to their small cell sizes. To resolve this critical issue, we built much larger supercells of $n \times n \times 2$ ($n$ = 4, 6, 8) containing up to 640 atoms. Unfortunately, for such large cells, exploring all possible spin (magnetic) configurations is not practically feasible. Thus, the configurations should be properly chosen to guarantee that our supercells reproduce the real charge distribution. Based on simple energy calculations, the spin states of the B-site ions are carefully deduced obeying the following rule: each nearest-neighbor (NN) Fe-Fe or Cr-Cr pair tends to have opposite (antiparallel) spins (see Table S2 in the Supporting Information for details of the analytic spin-interaction model). As a result, these large supercells possess lower K-S energies and larger band gaps than ordered BFCO, as shown in Fig. 1d; thus, they represent more realistic and disordered BFCOs observed in experiments.

The electron and hole densities of disordered BFCO in these supercells are shown in Fig. 3. Among the three different cell sizes ($n$ = 4, 6, 8) we explored, only the largest (i.e., 8×8×2 supercell) is shown. Comparing the CBM and VBM charge distributions in Fig. 3b-c, one can observe much more pronounced e-h separations than in the small cells. The CBM and VBM states are found to be clearly separated on a large scale at the Fe-rich and Cr-rich domains, respectively (Fig. 3b, 3c). We attribute this simulation result to the fact that, in disordered BFCO, structural symmetries existing in the ordered phase are broken, leading to the formation of locally BFO-like and $BiCrO_3$ (BCO)-like regions (Fig. 3a). This distinct separation phenomenon is found to be very general, irrespective of the supercell size or atomic configuration (see Fig. S4 in the Supporting Information for other supercell cases). Thus, the disordering of B-site cations in BFCO significantly enhances the e-h separation. The distinct separation in disordered BFCO reveals the origin of a prior experimental observation, that FPVs utilizing fully ordered BFCO perform inefficiently[16], and confirms the importance of disordered BFCO phases for efficient PV responses. Thus, charge separation can serve as the main reason for the phenomenal photocurrents in BFCO-based FPVs.

We discovered that the disordering of B-site cations not only enhances e-h separation but also greatly influences the charge-transport properties in BFCO. Using DFT, we calculated the charge-transport properties, namely, the effective masses of the photoexcited electron and hole. In Fig. 4, the effective masses in the ordered and disordered structures ($d_1$-$d_9$) are compared. Note that A-AFM ordering was not considered due to its relative instability (Fig. 1d). A prominent feature observed in Fig. 4 is that B-site ion disordering significantly reduces the effective mass of the electron ($m_e^*$). The mass of ordered BFCO is extremely large (47.3 $m_0$ for C-AFM and 44.3 $m_0$ for G-AFM). However, for most of the disordered BFCO cases (16 out of total 18 test samples), the electron becomes much lighter, showing an effective mass between 1.1 and 10.3 $m_0$. Only two cases ($d_3$ and $d_6$ with G-AFM ordering) deviate from the general trend and exhibit large $m_e^*$ values (43.9 $m_0$ and 34.8 $m_0$, respectively) comparable to that of ordered BFCO. In real BFCO films, all of these 18 disordered structures are likely to exist owing to their low and similar energies; thus, electrons would be on average much more mobile in the disordered domains (average mass: 8.2 $m_0$) than in the ordered domains (average mass: 45.8 $m_0$). The impact of B-site ion disordering on the effective mass of the hole ($m_h^*$) is relatively less dramatic. The effective masses of the holes are found to be quite small, in the range of 1.0 to 4.0 $m_0$, for both ordered and disordered geometries. See the Supporting Information for detailed orbital analysis, which explains the different quantities and behaviors of the electron and hole masses.

With the knowledge accumulated from the study of BFCO, we can provide a design strategy for next-generation FPVs. We explored 44 additional oxides with the aim to find novel light-absorbing materials that can offer efficient e-h separation for application in FPVs. The tested materials are all Bi-based double-perovskite structured oxides, i.e., $Bi_2(B_+,B_-)O_6$, where ($B_+,B_-$) represents possible combinations of transition-metal B-site cations (Fig. 5a). For each ($B_+,B_-$) combination, 6 phases can exist: $Fm\overline{3}m$, $R3$, $P4/mnc$, $I4/m$, $P2_1/n$, and $I2/m$[28]. The energies of all 6 phases were compared, and only the most stable phases were considered in the subsequent screening process. We calculated the degree of e-h separation, $R$ (defined in the Methods section), in each material. Systems with a negative $R$ value offer efficient spatial charge separation onto $B_+$ and $B_-$ cation sites. Out of 45 test sets, 10 ($B_+,B_-$) combinations showed negative $R$ values. We list these in the order from strong to weak separation: (Ti,V), (V,Cu), (Ti,Mn), (V,Mn), (Ti,Fe), (Cr,Ti), (Fe,Cr), (Co,Cr), (Mn,Co), and (Fe,V). Briefly note that, for these 10 ($B_+,B_-$) systems, the CBM and VBM states are spatially separated onto $B_+$ and $B_-$ sites, respectively. Our results indicate that 6 of these systems show better e-h separation than BFCO and could potentially serve as record-breaking active-layer materials in FPVs.

Along with the degree of charge separation, we also computed the $E_g$ values of the 45 test materials, as strong light absorption is paramount to realizing the potential of FPVs. The 10 aforementioned materials all exhibit much smaller $E_g$ values, ranging from 0.51 to 2.16 eV, than the prototypical FPV material $BiFeO_3$ (2.67 eV[29,30]); thus, much improved solar absorptions are expected. Note, however, that the $E_g$ values predicted using the DFT+$U$ scheme ($U$ = 4 eV and $J$ = 0.8 eV applied to the valence $d$-orbitals of both $B_+$ and $B_-$ ions) may differ from their experimental counterparts; comparisons are not possible at the present stage due to the absence of experimental data.

We also investigated the thermodynamic stability of the compounds, as there is no point in predicting structures that, at the very least, cannot be realized experimentally. Fig. 5b provides the formation energy of each compound, $Bi_2(B_+,B_-)O_6$, with respect to decomposition into $BiB_+O_3$ and $BiB_-O_3$. The suggested candidate materials mostly have negative formation energies, indicating they have the potential to be experimentally synthesized. In particular, $Bi_2(Ti,Fe)O_6$ is a great material for future work, as Ti-doped $BiFeO_3$ films have already been synthesized by several groups but not yet implemented in PV devices[31-33]; FPVs may immediately benefit from the strong e-h separations in this material.

Double-perovskite BFCO is undoubtedly the best performing active-layer material for FPVs and has indeed made FPVs much more competitive with conventional devices. The excellent performance was previously attributed primarily to the reduced band gap and resultant increase in solar absorption; however, our quantitative analysis reveals that this factor is not the primary reason. Based on electronic structure calculations of BFCO, we clarify that the photoexcited electron and hole states are spatially separated onto the Fe and Cr sites, respectively. Such e-h separation becomes much more pronounced in disordered phases, which is critical to understanding the extremely high photocurrents observed in BFCO-based FPV devices. In addition to providing an understanding of BFCO materials, the present work further suggests the design of next-generation FPV materials that can outperform BFCO. As a result of exploring 44 additional Bi-based double-perovskite oxides, we propose 9 novel light-absorbing materials that can offer strong e-h separation and a desired band gap for application in FPVs.

## Computational Methods

**First-principles calculations.** DFT calculations were performed using the plane-wave basis VASP code[33,34] with an energy cutoff of 500 eV. The projector-augmented-wave (PAW) method was adopted to describe the potential of the ionic cores. The geometry was fully relaxed until the maximum Hellmann-Feynman forces are less than 0.01 eV Å$^{-1}$. We employed the generalized gradient approximation (GGA) plus the Hubbard $U$ method[35,36] with the Perdew-Becke-Ernzerhof (PBE) exchange-correlation functional[37]. For all the BFCO oxides tested in the present work, we adopted $U$ = 4 eV and $J$ = 0.8 eV for both Fe and Cr 3$d$ orbitals, as described in previous reports, in order to accurately describe the band structures of BFCO[17,18]. For the additional 44 double-perovskite structured oxides, Bi$_2$(B$_+$,B$_-$)O$_6$, in Fig. 5, we applied the same Hubbard $U$ parameters, $U$ = 4 eV and $J$ = 0.8 eV, to the valence $d$-orbitals of both B$_+$ and B$_-$ ions. We used different $k$-point meshes depending on the cell sizes. For the simulation cells in Figs. 1, 2, 4 and 5, Brillouin-zone integrations were performed using Monkhorst-Pack $k$-point samplings of 3×3×3 for structural relaxation after the convergence tests and 9×9×9 for band structures and charge densities. For the 8×8×2 supercell shown in Fig. 3, a $k$-point mesh of 1×1×4 was chosen for both geometry relaxation and charge density calculations.

**Effective mass calculations.** The effective mass ($m^*$) was calculated according to

$$m^* = \pm \hbar^2 \left(\frac{d^2E}{dk^2}\right)^{-1},$$

where $\hbar$ is the reduced Planck constant and $E$ is the energy of a band as a function of the wave vector $k$. The fitting range is $|k|$ < 0.05 Å$^{-1}$.

**Quantification of the degree of e-h separation.** The degree of e-h separation was estimated by introducing a DOS-related factor ($R$), which is defined as

$$R = \left(\frac{S_{B_+}-S_{B_-}}{S_{B_+}+S_{B_-}}\right)_{CBM} \times \left(\frac{S_{B_+}-S_{B_-}}{S_{B_+}+S_{B_-}}\right)_{VBM},$$

where $S_{B_+}$ is the integrated DOS value of the $B_+$ cation, while $S_{B_-}$ denotes the integrated DOS value of the $B_-$ cation in the immediate vicinity of either the CBM or VBM. The integration was performed over an energy range of 0.05 eV (see Fig. S5 in the Supporting Information). In principle, the $R$ value can range from −1 to +1. Systems with negative $R$ values offer efficient spatial charge separation. A more negative $R$ indicates stronger charge separation. $R = -1$ represents *complete* e-h separation onto two distinct *B*-cation sites.

## Acknowledgements

This work was supported by the National Research Foundation (NRF Grant No. 2016R 1D1A1B 03933253) of Korea, and institutional project of KIST (Project No. 2E26130). Authors are grateful to Dr. Sang Soo Han, Dr. Seungchul Kim, and Dr. Sungjin Pai at KIST for their constructive discussions.

# Figures

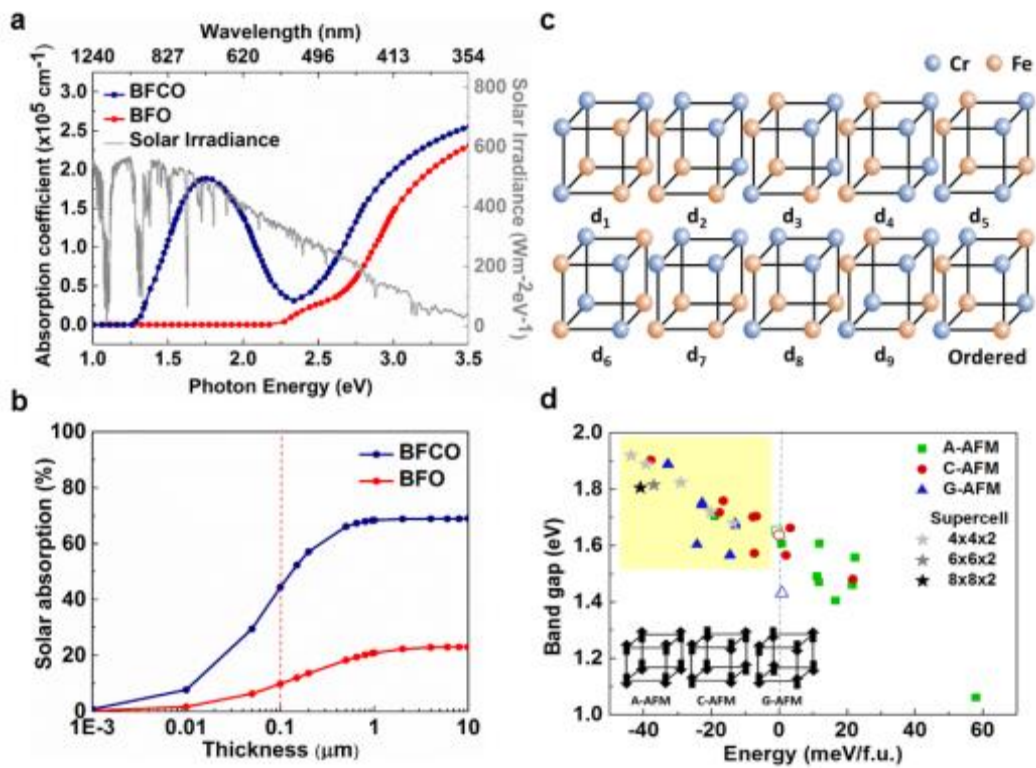

**Figure 1. Solar absorption, ion configurations, and band gaps of BFCO materials. a**, Absorption coefficients [reproduced from $(\alpha E)^2$ vs. $E$ plots provided in ref. 16] plotted as a function of the photon energy. The gray line represents the AM 1.5G (100 mW/cm$^2$) solar irradiance. **b**, Thickness-dependent solar absorption (%) curves of the BFO and BFCO films. **c**, Schematic diagrams of 10 possible B-site cation configurations in 40-atom BFCO unit cells, including the ordered case. The 9 disordered structures are labeled from $d_1$ to $d_9$. **d**, Band gap *vs*. K-S energy plot for BFCO materials with all atomic and magnetic configurations. Ordered structures are indicated by open symbols (near energy = 0 eV). Supercells ($n$ = 4, 6, 8) are denoted by gray stars. Here, the energy in the abscissa is referenced to the K-S energy of C-AFM ordered BFCO. Structures in the yellow-shaded region well represent the disordered BFCO phases observed in experiments.

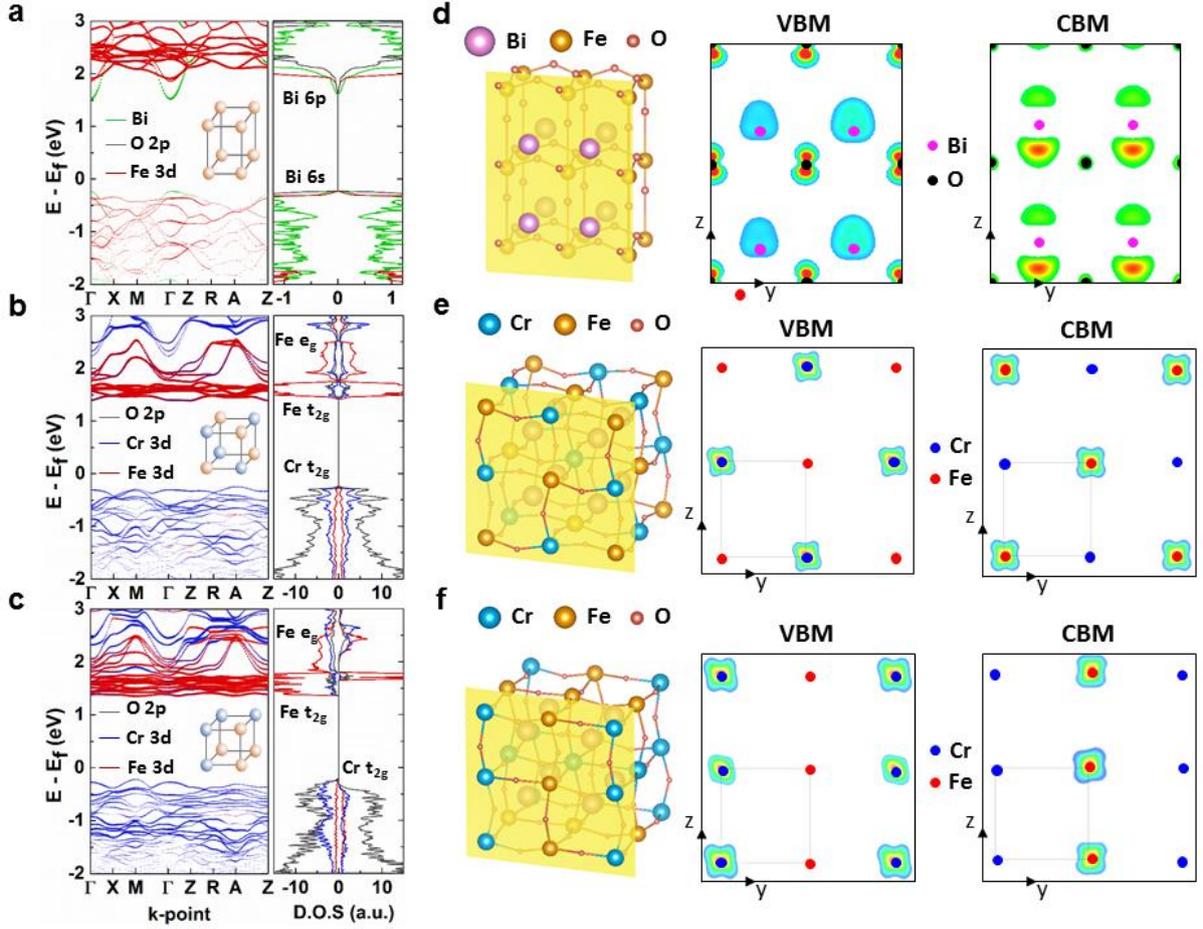

**Figure 2. Electronic structures of BFO and BFCO. a-c**, Calculated orbital-resolved band structures and the corresponding DOS of BFO with C-AFM ordering (**a**), ordered BFCO with C-AFM ordering (**b**), and "$d_1$" structure (disordered) BFCO with C-AFM ordering (**c**). In the band structure plots, the red, blue, and green colors indicate the contributions of the Fe 3$d$, Cr 3$d$, and Bi orbitals, respectively. Similarly, in the DOS plots, the red, blue, green, and gray lines represent the Fe 3$d$, Cr 3$d$, Bi, and O 2$p$ states, respectively. **d-f**, 2×2×2 supercell structure (left), and 2D cross-sections of the square of wavefunctions (i.e., partial charge densities) of the VBM (middle) and CBM (right): BFO with C-AFM ordering (**d**), ordered BFCO with C-AFM ordering (**e**), and "$d_1$" BFCO with C-AFM ordering (**f**). Note that, in Fig. 2**e** and 2**f**, oxygen atoms and their contributions are not shown for clarity, as they are out of the cross-section planes. Three-dimensional plots of the partial charge densities are available in Fig. S3 in the Supporting Information.

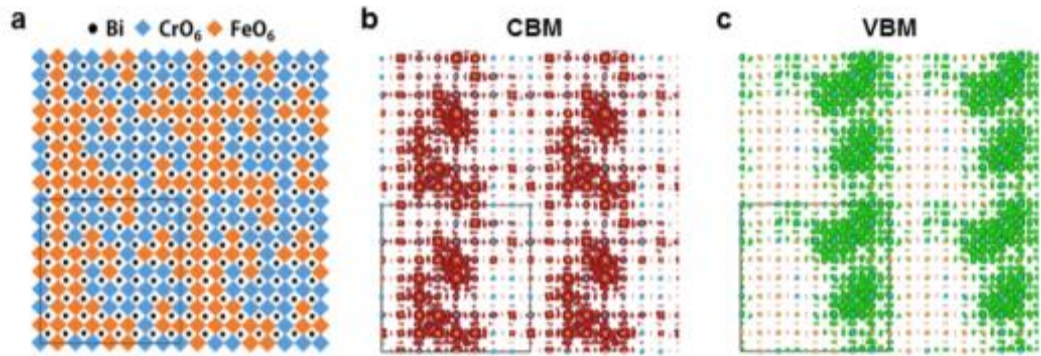

**Figure 3. Distinct e-h separation in disordered BFCO on a large scale. a**, Schematic of the tested 8×8×2 supercell. **b-c**, Partial charge densities of the CBM (**b**, in the red cloud) and VBM (**c**, in the green cloud) states, showing that the CBM and VBM states are spatially separated on the Fe-rich and Cr-rich domains, respectively.

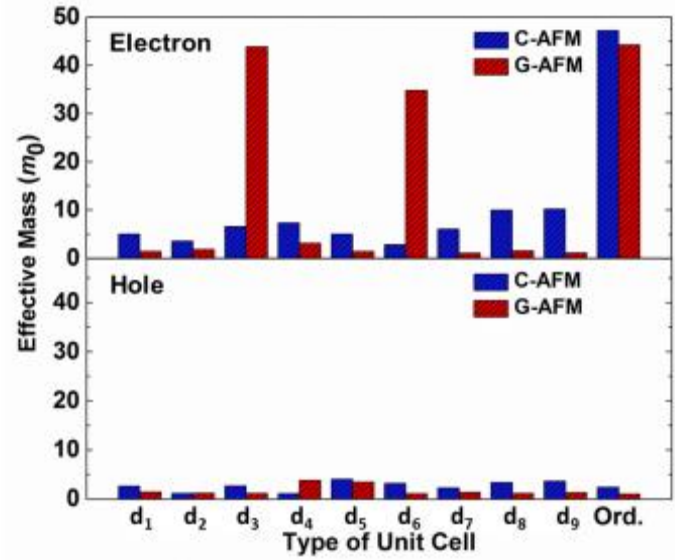

**Figure 4. B-site ion configuration-dependent charge-transport properties.** Calculated electron (upper panel) and hole (lower panel) effective masses in the direction of polarization (i.e., $z$-axis) for ordered BFCO and disordered BFCO ($d_1$-$d_9$).

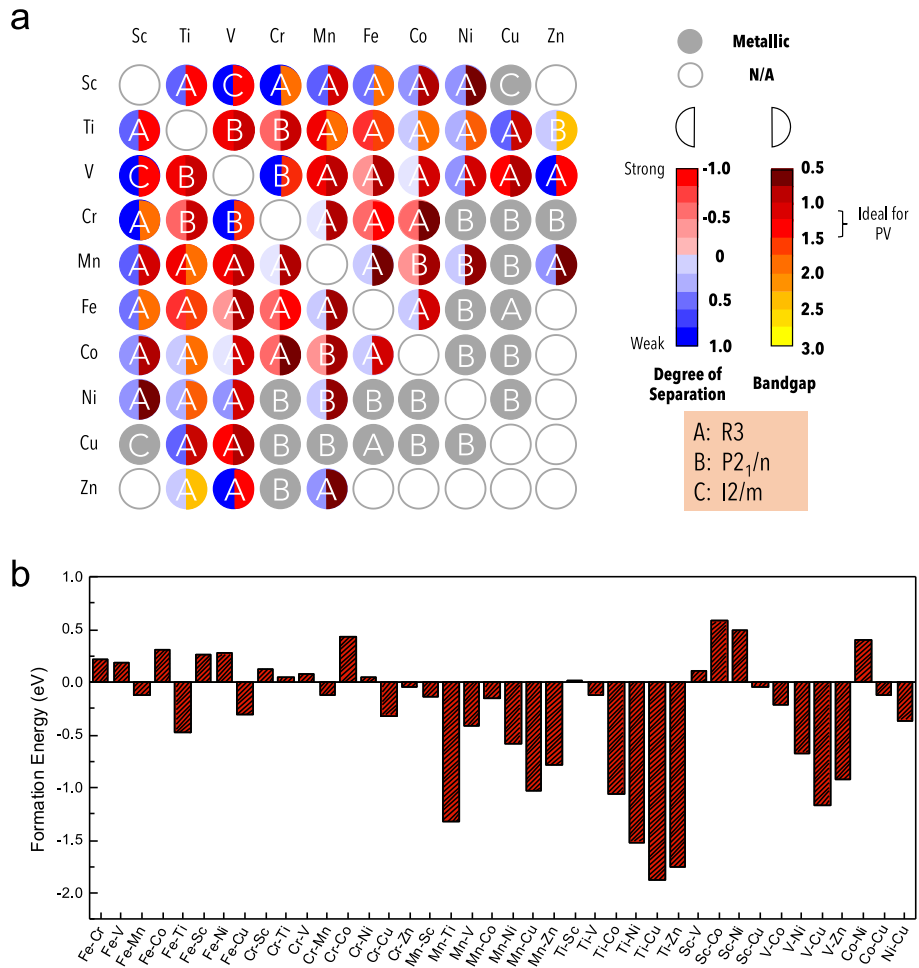

**Figure 5. In search of candidate materials with efficient FPV responses.** The 45 tested materials are all $Bi_2(B_+,B_-)O_6$-type double-perovskite structured oxides, where $(B_+,B_-)$ represents the possible combinations of transition-metal B-site cations. For fast screening, we reasonably assumed G-AFM ordered geometries. **a**. The degree of e-h separation (left semi-circle) and band gap (right semi-circle) of these oxides are shown with the color mapping. The letter (A, B, C) inside each circle represents the corresponding ground-state phase. In reality, the open gray-circle cases cannot exist due to oxidation-state mismatch. The closed gray circles represent metallic samples. **b**. Formation energies of the $Bi_2(B_+,B_-)O_6$ compounds with respect to decomposition into $BiB_+O_3$ and $BiB_-O_3$. Most of the double perovskites have lower formation energies than BFCO.

# Supplementary Information

- **Photovoltaic performance of the $BiFeO_3$-film based PVs**

|  | Up | Down |
|---|---|---|
| $J_{sc}$ (mA/cm$^2$) | 0.01155 | −0.0104 |
| $V_{oc}$ (V) | −0.38 | 0.30 |
| F. F. (%) | 28.63 | 29.87 |
| PCE (%) | 0.00126 | 0.00093 |

**Table S1. Photovoltaic performance of the $BiFeO_3$-film-based solar cell with upward and downward poling measured under AM 1.5G 100 mW/cm$^2$ simulated sunlight.** The solar cell has the following heterojunction structure: ITO/$BiFeO_3$/$SrRuO_3$/$SrTiO_3$. The [001]-oriented $BiFeO_3$ thin film used here is 100-nm-thick.

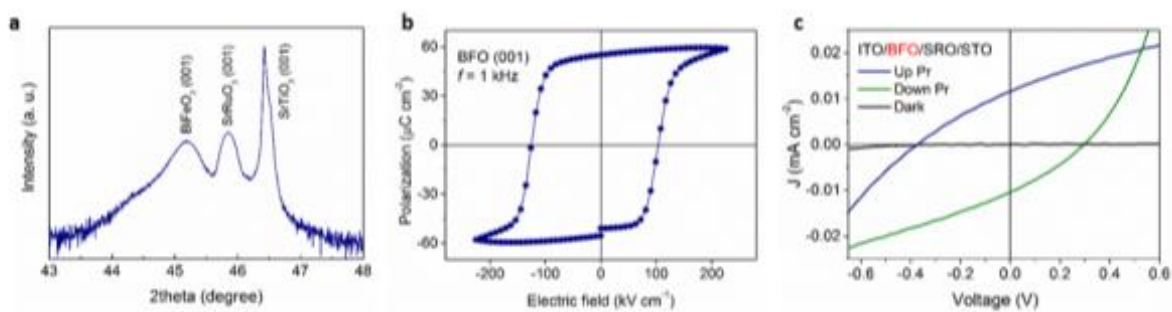

**Fig. S1. Structural, ferroelectric, and photovoltaic performance data for the 100-nm-thick $BiFeO_3$ thin films grown on a $SrRuO_3$/$SrTiO_3$ substrate. a**, Theta-2theta ($\theta - 2\theta$) X-ray diffraction (XRD) patterns of the [001]-oriented $BiFeO_3$ films prepared by pulsed laser deposition method. **b**, Polarization-electric field (*P-E*) hysteresis curve obtained at 300 K. **c**, J-V response curves under dark condition (black with negligibly small photocurrents) and under AM 1.5G 100mW/cm$^2$ illumination with upward poling (blue) and downward poling (green).

- **Analytic energy modeling for large size supercells (relevant to Fig. 1 in the main text)**

The B-site ion total energy ($E_{tot}$) can be expressed using the following equation:

$$E_{tot} = N_{Fe}U_{Fe} + N_{Cr}U_{Cr} + \sum_{(i,j)} N_{(i,j)}\varepsilon_{(i,j)}$$

where $N_{Fe}$ ($N_{Cr}$) is the number of Fe (Cr) ions in a unit cell, whereas $U_{Fe}$ and $U_{Cr}$ denote the on-site energy terms for Fe and Cr ions, respectively. Here, (i,j) counts the nearest-neighbor (NN) B-site cation pairs. We have considered six interaction terms, i.e., (Fe↑,Fe↑), (Fe↑,Fe↓), (Cr↑,Cr↑), (Cr↑,Cr↓), (Fe↑,Cr↑), and (Fe↑,Cr↓). $N_{(i,j)}$ and $\varepsilon_{(i,j)}$ are the number of (i,j) pairs in a unit cell and the NN interaction energy of (i,j) pair, respectively.

|  | spin | $N_{Fe}$ | $N_{Cr}$ | $N_{(Fe\uparrow,Fe\uparrow)}$ | $N_{(Fe\uparrow,Fe\downarrow)}$ | $N_{(Cr\uparrow,Cr\uparrow)}$ | $N_{(Cr\uparrow,Cr\downarrow)}$ | $N_{(Fe\uparrow,Cr\uparrow)}$ | $N_{(Fe\uparrow,Cr\downarrow)}$ | Energy (eV) |
|---|---|---|---|---|---|---|---|---|---|---|
| $d_1$ | A | 4 | 4 | 4 | 2 | 4 | 2 | 8 | 4 | -32.9157 |
|  | C | 4 | 4 | 2 | 4 | 2 | 4 | 4 | 8 | -32.9347 |
|  | G | 4 | 4 | 0 | 6 | 0 | 6 | 0 | 12 | -32.9501 |
| $d_2$ | A | 4 | 4 | 4 | 2 | 4 | 2 | 8 | 4 | -32.9162 |
|  | C | 4 | 4 | 2 | 4 | 2 | 4 | 4 | 8 | -32.9350 |
|  | G | 4 | 4 | 0 | 6 | 0 | 6 | 0 | 12 | -32.9516 |
| $d_3$ | A | 4 | 4 | 2 | 2 | 2 | 2 | 12 | 4 | -32.9266 |
|  | C | 4 | 4 | 2 | 2 | 2 | 2 | 4 | 12 | -32.9254 |
|  | G | 4 | 4 | 0 | 4 | 0 | 4 | 0 | 16 | -32.9418 |
| $d_4$ | A | 4 | 4 | 4 | 4 | 4 | 4 | 8 | 0 | -32.9108 |
|  | C | 4 | 4 | 4 | 4 | 4 | 4 | 0 | 8 | -32.9241 |
|  | G | 4 | 4 | 0 | 8 | 0 | 8 | 0 | 8 | -32.9600 |
| $d_5$ | A | 4 | 4 | 4 | 2 | 4 | 2 | 8 | 4 | -32.9155 |
|  | C | 4 | 4 | 2 | 4 | 2 | 4 | 4 | 8 | -32.9341 |
|  | G | 4 | 4 | 0 | 6 | 0 | 6 | 0 | 12 | -32.9501 |
| $d_6$ | A | 4 | 4 | 0 | 4 | 0 | 4 | 16 | 0 | -32.9463 |
|  | C | 4 | 4 | 4 | 0 | 4 | 0 | 0 | 16 | -32.9056 |
|  | G | 4 | 4 | 0 | 4 | 0 | 4 | 0 | 16 | -32.9402 |
| $d_7$ | A | 4 | 4 | 8 | 0 | 8 | 0 | 0 | 8 | -32.8693 |
|  | C | 4 | 4 | 0 | 8 | 0 | 8 | 8 | 0 | -32.9652 |
|  | G | 4 | 4 | 0 | 8 | 0 | 8 | 0 | 8 | -32.9603 |
| $d_8$ | A | 4 | 4 | 4 | 0 | 4 | 0 | 8 | 8 | -32.9058 |
|  | C | 4 | 4 | 0 | 4 | 0 | 4 | 8 | 8 | -32.9450 |
|  | G | 4 | 4 | 0 | 4 | 0 | 4 | 0 | 16 | -32.9419 |
| $d_9$ | A | 4 | 4 | 4 | 0 | 4 | 0 | 8 | 8 | -32.9050 |
|  | C | 4 | 4 | 0 | 4 | 0 | 4 | 8 | 8 | -32.9438 |
|  | G | 4 | 4 | 0 | 4 | 0 | 4 | 0 | 16 | -32.9404 |
| Ordered | A | 4 | 4 | 0 | 0 | 0 | 0 | 16 | 8 | -32.9280 |
|  | C | 4 | 4 | 0 | 0 | 0 | 0 | 8 | 16 | -32.9273 |
|  | G | 4 | 4 | 0 | 0 | 0 | 0 | 0 | 24 | -32.9266 |

**Table. S2. Numbers of Fe, Cr ions and NN B-site ion pairs, and K-S energies for various B-site ion configuration.**

Using the quasi-newton methods (implemented in the software MATHEMATICA), we have obtained the fitted on-site and NN interaction energy terms. These are summarized in the below:

(i)   $U_{Fe} = = -4.02875$ (eV)   and   $U_{Cr} = = -4.02875$ (eV),

(ii)  $\varepsilon_{(Fe\uparrow,Fe\uparrow)} = -25.97$ (meV)   *versus*   $\varepsilon_{(Fe\uparrow,Fe\downarrow)} = -31.13$ (meV),

(iii) $\varepsilon_{(Cr\uparrow,Cr\uparrow)} = -25.97$ (meV)   *versus*   $\varepsilon_{(Cr\uparrow,Cr\downarrow)} = -31.13$ (meV),

(iv) $\varepsilon_{(Fe\uparrow,Cr\uparrow)} = -29.29$ (meV)   *versus*   $\varepsilon_{(Fe\uparrow,Cr\downarrow)} = -29.00$ (meV).

From the comparisons of the fitted interaction energy values, we have drawn several conclusions regarding the NN spin interactions in BFCO. These can be summarized as:

"In order to **minimize** the B-site ion interaction energy, both the **NN Fe-Fe pairs** and the **NN Cr-Cr pairs** should have **antiparallel** spin states (or opposite spins) with $\varepsilon_{(i\uparrow,j\downarrow)}$ of $-31.13$ meV per pair. On the other hand, for the dissimilar **NN Fe-Cr pairs**, antiparallel spin states do not have any noticeable preference over parallel spin states. Notice that the ordered BFCO is exclusively characterized by this dissimilar NN Fe-Cr pairs. In addition to these rules, the above computational results indicate that the NN Fe-Fe or NN Cr-Cr antiparallel spin pair is energetically more favorable than the NN Fe-Cr pair (either parallel or antiparallel spins), namely, $\varepsilon_{(i\uparrow,j\downarrow)}$ of $-31.13$ meV *vs.* $\varepsilon_{(i,i)}$ of ~ $-29$ meV. This further suggests the relative stability of disordered BFCOs over the ordered BFCO in terms of the NN interaction energy."

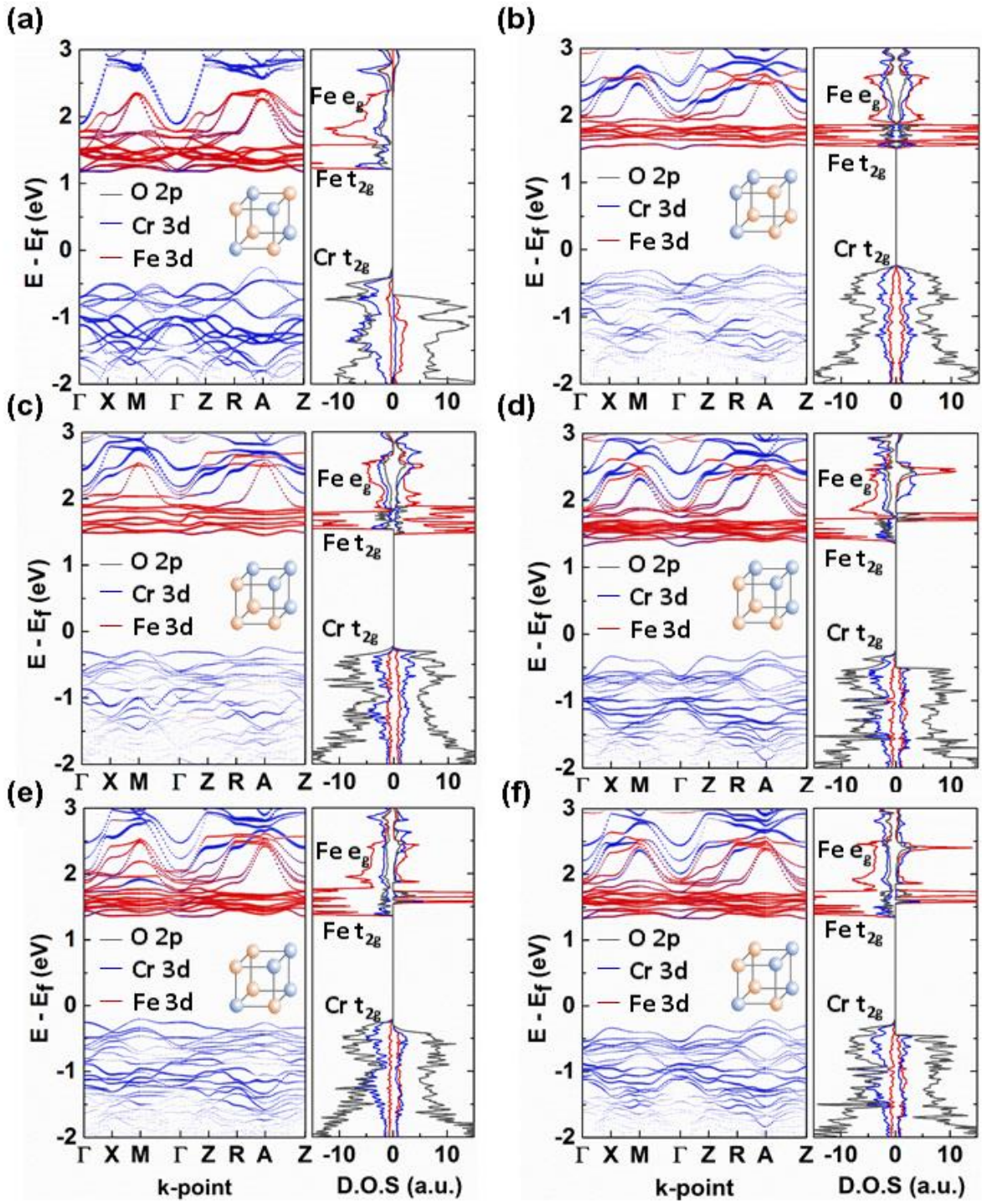

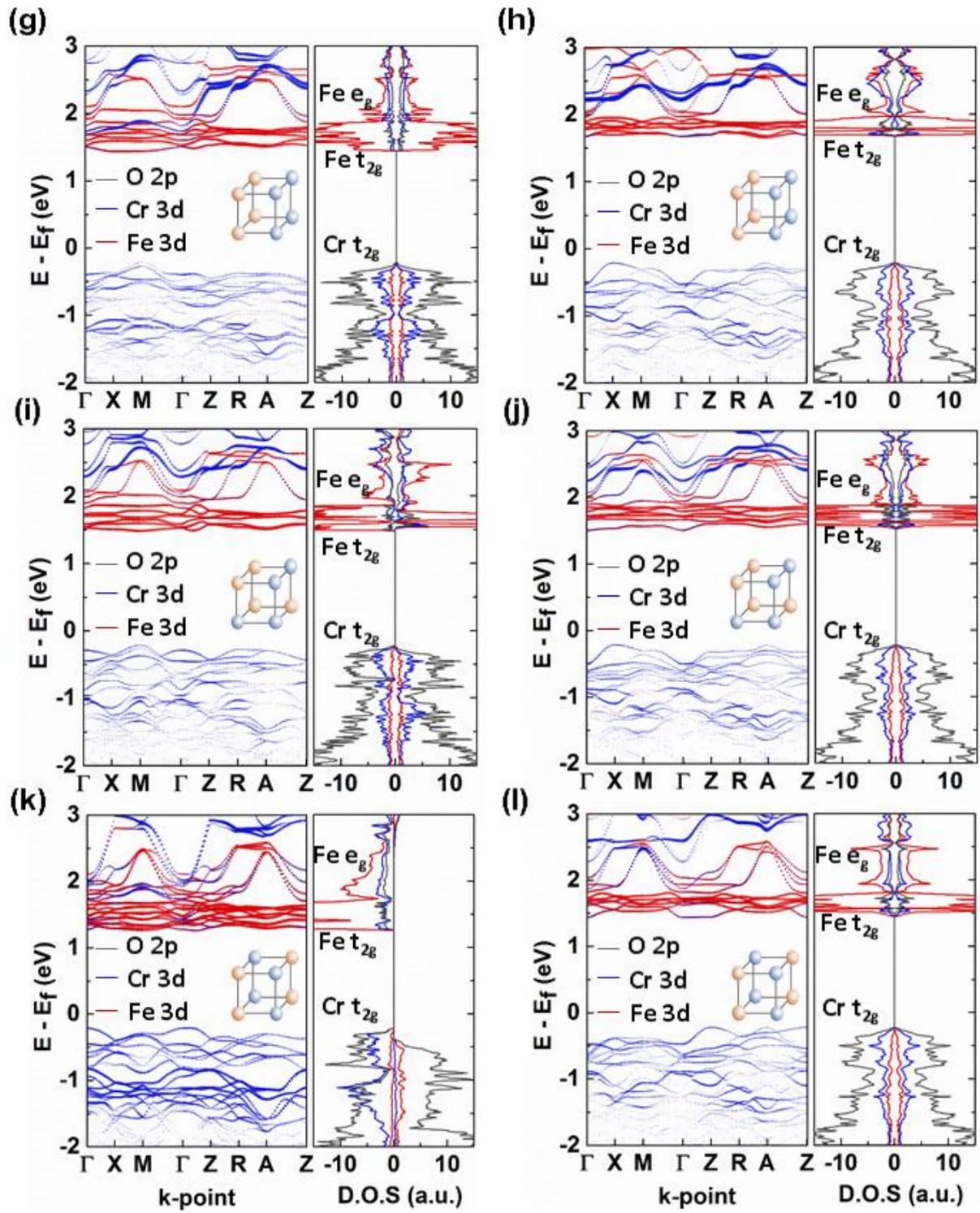

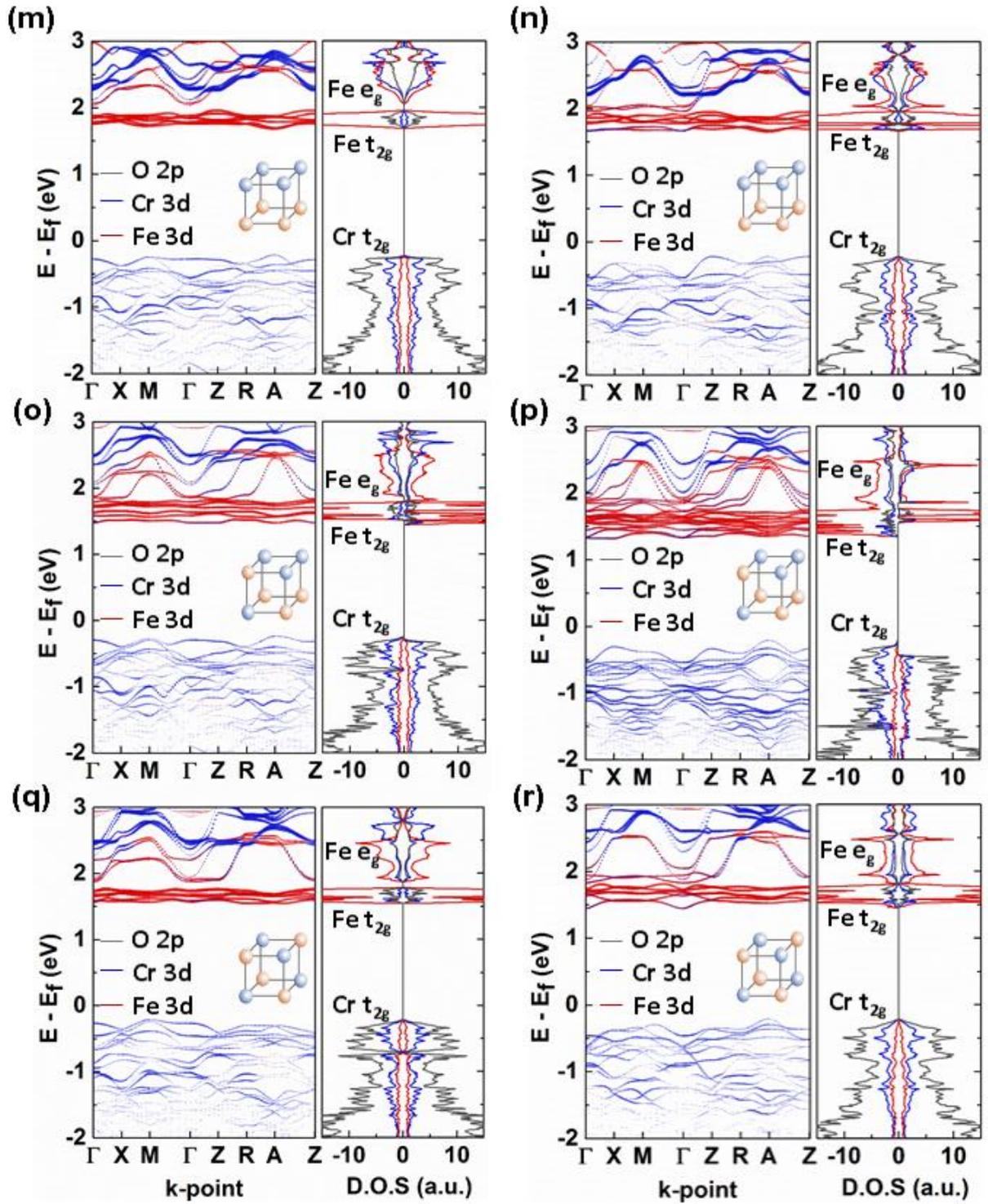

**Fig. S2. Orbital resolved band structure and DOS for all atomic and magnetic orderings.**
**a**, Ordered in G-AFM, **b**, $d_1$ in G-AFM, **c**, $d_2$ in C-AFM, **d**, $d_2$ in G-AFM, **e**, $d_3$ in C-AFM, **f**, $d_3$ in G-AFM, **g**, $d_4$ in C-AFM, **h**, $d_4$ in G-AFM, **i**, $d_5$ in C-AFM, **j**, $d_5$ in G-AFM, **k**, $d_6$ in C-AFM, **l**, $d_6$ in G-AFM, **m**, $d_7$ in C-AFM, **n**, $d_7$ in G-AFM, **o**, $d_8$ in C-AFM, **p**, $d_8$ in G-AFM, q, $d_9$ in C-AFM, **r**, $d_9$ in G-AFM.

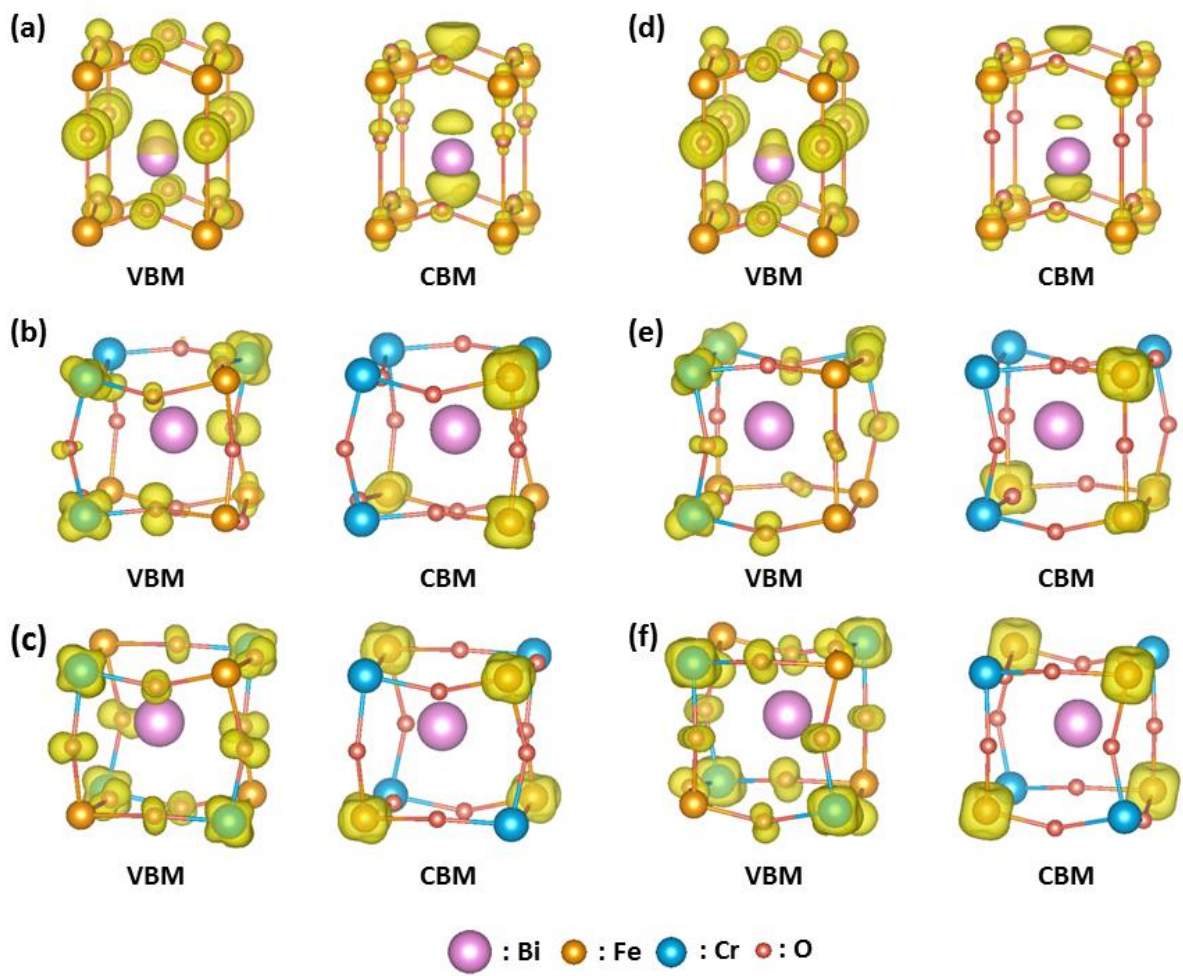

**Fig. S3. Three-dimensional plots of electron and hole densities of BFO and BFCO. a-c,** Partial charge densities (VBM and CBM) of BFO in C-AFM (**a**), "d$_1$" BFCO in C-AFM (**b**), ordered BFCO in C-AFM (**c**), and **d-f**, those of BFO in G-AFM (**d**), "d$_1$" BFCO in G-AFM (**e**), ordered BFCO in G-AFM (**f**).

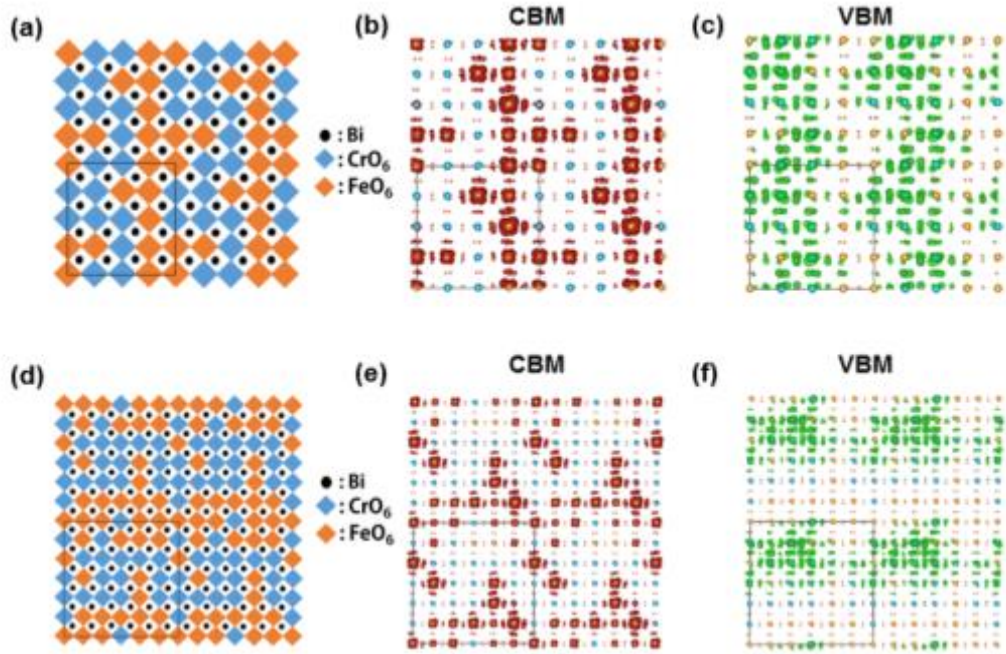

**Fig. S4. Distinct charge separation in disordered BFCOs in different scales. a-c,** a schematic diagram of the 4×4×2 supercell (**a**) and corresponding CBM (**b**) and VBM (**c**) charge densities, **d-f**, a schematic diagram of the 6×6×2 supercell (**d**) and corresponding CBM (**e**) and VBM (**f**) charge densities. In both cells, CBM and VBM states clearly show e-h separations in disordered BFCO phases.

- **Definition of the degree of charge separation**

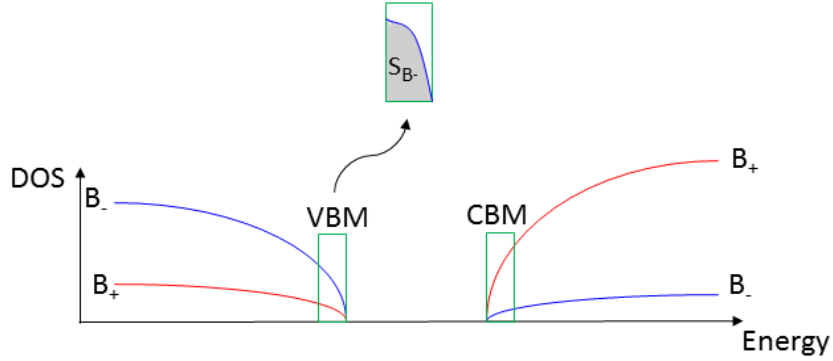

**Fig. S5. A schematic diagram of the density of states in the vicinity of the band gap for the Bi-based double-perovskite, Bi(B$_+$,B$_-$)$_{0.5}$O$_3$.**

To quantitatively compare the degree of e-h charge separation in the double-perovskite materials, we define the degree of charge separation ($R$) as

$$R = \left(\frac{S_{B_+} - S_{B_-}}{S_{B_+} + S_{B_-}}\right)_{CBM} \times \left(\frac{S_{B_+} - S_{B_-}}{S_{B_+} + S_{B_-}}\right)_{VBM},$$

where $S_{B_+}$ is the integrated DOS value of the $B_+$ cation while $S_{B_-}$ denotes the integrated DOS value of the $B_-$ cation in the immediate vicinity of either CBM or VBM. The integration was performed over the energy range of 0.05 eV. In principle, the $R$-value ranges from −1 to +1. Systems of negative $R$-values can offer efficient spatial charge separation. The more negative $R$ is, the stronger separation is. $R = -1$ represents the *complete* e-h separation onto two distinct $B$-cation sites.

- **The degree of charge separation for 45 Bi$_2$(B$_+$,B$_-$)O$_6$-type double-perovskite structured oxides**

|    | Sc   | Ti    | V     | Cr    | Mn    | Fe    | Co    | Ni   | Cu    | Zn   |
|----|------|-------|-------|-------|-------|-------|-------|------|-------|------|
| Sc |      | 0.56  | 0.90  | 0.89  | 0.78  | 0.71  | 0.63  | 0.62 | M     | N/A  |
| Ti | 0.56 |       | -0.89 | -0.67 | -0.87 | -0.82 | 0.10  | 0.27 | 0.55  | 0.05 |
| V  | 0.90 | -0.89 |       | 0.90  | -0.86 | -0.41 | 0.01  | 0.45 | -0.87 | 0.93 |
| Cr | 0.89 | -0.67 | 0.90  |       | 0.03  | -0.57 | -0.54 | M    | M     | M    |
| Mn | 0.78 | -0.87 | -0.86 | 0.03  |       | 0.07  | -0.43 | 0.03 | M     | 0.63 |
| Fe | 0.71 | -0.82 | -0.41 | -0.57 | 0.07  |       | 0.05  | M    | M     | N/A  |
| Co | 0.63 | 0.10  | 0.01  | -0.54 | -0.43 | 0.05  |       | M    | M     | N/A  |
| Ni | 0.62 | 0.27  | 0.45  | M     | 0.03  | M     | M     |      | M     | N/A  |
| Cu | M    | 0.55  | -0.87 | M     | M     | M     | M     | M    |       | N/A  |
| Zn | N/A  | 0.05  | 0.93  | M     | 0.63  | N/A   | N/A   | N/A  | N/A   |      |

Table S2. Calculated values of the degree of charge separation (R) for 45 Bi$_2$(B$_+$,B$_-$)O$_6$-type double-perovskite structured oxides.

- **The calculated band gap for 45 $Bi_2(B_+,B_-)O_6$-type double-perovskite structured oxides**

|    | Sc   | Ti   | V    | Cr   | Mn   | Fe   | Co   | Ni   | Cu   | Zn   |
|----|------|------|------|------|------|------|------|------|------|------|
| Sc |      | 1.48 | 1.49 | 2.15 | 1.20 | 2.12 | 1.06 | 0.50 | M    | N/A  |
| Ti | 1.48 |      | 1.01 | 1.17 | 2.16 | 1.68 | 2.16 | 2.30 | 0.95 | 2.80 |
| V  | 1.49 | 1.01 |      | 1.37 | 1.27 | 1.03 | 1.21 | 1.28 | 0.99 | 1.48 |
| Cr | 2.15 | 1.17 | 1.37 |      | 1.09 | 1.45 | 0.51 | M    | M    | M    |
| Mn | 1.20 | 2.16 | 1.27 | 1.09 |      | 0.73 | 0.85 | 0.62 | M    | 0.72 |
| Fe | 2.12 | 1.68 | 1.03 | 1.45 | 0.73 |      | 1.06 | M    | M    | N/A  |
| Co | 1.06 | 2.16 | 1.21 | 0.51 | 0.85 | 1.06 |      | M    | M    | N/A  |
| Ni | 0.50 | 2.30 | 1.28 | M    | 0.62 | M    | M    |      | M    | N/A  |
| Cu | M    | 0.95 | 0.99 | M    | M    | M    | M    | M    |      | N/A  |
| Zn | N/A  | 2.80 | 1.48 | M    | 0.66 | N/A  | N/A  | N/A  | N/A  |      |

**Table S3. Calculated values of the band gap for 45 Bi-based double-perovskite oxides, $Bi_2(B_+,B_-)O_6$.**

| Configuration Number | Electron (C-AFM) | Hole (C-AFM) | Electron (G-AFM) | Hole (G-AFM) |
|---|---|---|---|---|
| $d_1$ | 5.07 | 2.67 | 1.45 | 1.41 |
| $d_2$ | 3.60 | 1.13 | 1.85 | 1.21 |
| $d_3$ | 6.62 | 2.73 | 43.88 | 1.17 |
| $d_4$ | 7.35 | 1.12 | 3.14 | 3.84 |
| $d_5$ | 5.07 | 4.07 | 1.49 | 3.52 |
| $d_6$ | 2.87 | 3.22 | 34.85 | 1.10 |
| $d_7$ | 6.09 | 2.34 | 1.12 | 1.36 (A) |
| | | | | 0.99 (Z) |
| $d_8$ | 10.00 | 3.41 | 1.57 | 1.16 |
| $d_9$ | 10.25 | 3.67 | 1.16 | 1.31 |
| Ordered | 47.26 | 2.45 | 44.26 | 1.03 |

**Table S4. Computed electron and hole effective masses of BFCO under C-AFM and G-AFM orderings for various B-site ion configurations expressed in unit of $m_0$.** The electron and hole effective masses were calculated at the CBM and VBM, respectively. Note that the $d_7$ structure in G-AFM can have two possible hole effective masses because, in this particular case, the VBMs exist at two distinct high-symmetry points, A and Z.

The different quantities and behaviors between the electron and hole effective mass can be understood by analyzing the orbital hybridizations near the CB and VB edges, respectively, since the effective mass value is highly dependent on the orbital hybridization. To start with the CBM state where Fe $3d$ orbitals dominate, the NN Fe $3d$-Fe $3d$ interactions are much weaker in the ordered BFCO structure due to the average longer Fe-Fe distance than *any* disordered BFCO (Fig. 2e-f). Thus, the electron's mass is substantially reduced in the disordered geometries. Unlike the CBM where O $2p$ contribution is negligible, the VBM state constitutes not only Cr $3d$ but also O $2p$ orbitals. As these O $2p$ orbitals act as bridges between the NN Cr $3d$ orbitals (i.e., Cr-O-Cr), the orbital hybridization is much stronger, leading to a smaller hole's effective mass.

- **REFERENCES**